\documentclass{kapedbk}

\usepackage[english]{babel}

\usepackage{amsmath,amsfonts,amssymb}
\usepackage{latexsym}
\usepackage{epsfig}
\usepackage{graphics}
\usepackage{graphicx}
\usepackage{rotating}

\let\footnote\savefootnote

\let\footnotetext\savefootnotetext

\let\footnoterule\savefootnoterule

\normallatexbib

\begin{document}
\setcounter{page}{239}
\sloppy

\articletitle[Self-Organized Control of Ir\-regular or Perturbed
Net\-work Traffic]{Self-Organized Control of\\ Ir\-regular or Perturbed\\
Net\-work Traffic}

\chaptitlerunninghead{Self-Organized Control of Ir\-regular or Perturbed
Net\-work Traffic}

\author{Dirk Helbing}
\affil{Dresden University of Technology, Andreas-Schubert-Str. 23,
01062 Dresden} \email{helbing@trafficforum.org} \vspace*{-10pt}

\author{Stefan L\"ammer}
\affil{Dresden University of Technology, Andreas-Schubert-Str. 23,
01062 Dresden} \email{traffic@stefanlaemmer.de} \vspace*{-10pt}

\author{Jean-Patrick Lebacque}
\affil{Institut National de Recherche sur les Transports er leur
S\'ecurit\'e (INRETS), \\ 2 av. G\'en\'eral Malleret-Joinville,
F-94114 ARCUEIL CEDEX, France} \email{lebacque@inrets.fr}
\vspace*{-10pt}

\begin{abstract}
We present a fluid-dynamic model for the simulation of urban traffic networks with
road sections of different lengths and capacities. The model allows one to
efficiently simulate the transitions between free and congested traffic,
taking into account congestion-responsive traffic assignment and adaptive traffic control. We observe dynamic
traffic patterns which significantly depend on the respective network topology. Synchronization
is only one interesting example and implies the emergence of green waves.
In this connection, we will discuss adaptive strategies of traffic light
control which can considerably improve throughputs and travel times, using self-organization
principles based on local interactions between vehicles and traffic lights.
Similar adaptive control principles can be applied to other queueing
networks such as production systems. In fact, we suggest to turn push operation of
traffic systems into pull operation: By removing vehicles as fast as possible from the
network, queuing effects can be most efficiently avoided.
The proposed control concept can utilize the cheap sensor
technologies available in the future and leads to reasonable operation modes. It is flexible,
adaptive, robust, and decentralized rather than based on precalculated
signal plans and a vulnerable traffic control center.
\end{abstract}


\section{Introduction}

Traffic control in networks has a long history. Early efforts have
aimed at synchronizing traffic signals along a one-way, then a two-way
arterial. There is still potential for improvement in this direction,
as is attested by some recent research efforts \cite{aut-gar99} or
prompted by the development of new theoretical tools \cite{aut-lot02,aut-man01}.
Synchronization of traffic along arterials results in so-called
green-waves, the aim of which is simply to ensure that traffic flows
smoothly along main streets. Expected benefits of green waves
are reduced fuel consumption and travel times.

The green-wave approach can be generalized to networks, yielding
pre-calculated signal control schemes, such as TRANSYT \cite{aut-rob97}.
In principle such schemes are completely coercive: they force the
traffic flow to comply with pre-calculated patterns, optimizing such
criteria as the total travel time spent. Since traffic demand varies, the
need for some responsiveness of the signal control was felt very
soon. The SCOOT system \cite{aut-rob91}, an outgrowth of TRANSYT,
allows for smooth change in the signal settings in response to changes
in the traffic demand.

Among the strategies making use of precalculated controls, let us
mention SCATS \cite{aut-sims79,aut-lin04}, which relies on a
library of controls (green durations,
offsets, ...) according to traffic conditions. Even the optimization
criterion depends on the traffic state. The system might, at night,
minimize the number of stops, maximize throughput at day time under normal conditions,
and aim at postponing the onset of congestion under heavy traffic conditions.

More recent developments stress greater adaptability. For instance
UTOPIA \cite{aut-tar89} combines a regional control based on prediction
of traffic flow through the main network arteries with the action of local
intersection controllers. The regional control simply serves as a
reference for local control.

OPAC \cite{aut-gart90} optimizes queues in accordance with the
``store-and-forward'' concept \cite{Papa91}, based on
dynamic programming, with a rolling horizon. OPAC is fundamentally
designed to manage intersections but extends to networks.

Even more decentralized and demand-responsive at a very local level,
PRODYN \cite{aut-hen90} optimizes traffic at intersections by switching
traffic lights on a traffic-actuated basis. Optimality is achieved
through the dynamic programming technique. PRODYN also tries to coordinate
neighboring intersections.

A further development includes dynamic assignment into the calculation of
optimal traffic light settings as well as non-mandatory management
schemes (user information). METACOR \cite{aut-ell94}, based on an optimal
control strategy with a rolling horizon, is a good example of this
approach. In the same line of approach, TUC \cite{aut-dia03} displays
two innovative features:
\begin{itemize}
\item[1.] a reference strategy is calculated for the network (for a given
  situation),
\item[2.] a filter is included into the algorithm which calculates the
  commands. The aim of the filter is to detect and adjust deviations from
  the nominal traffic situation, and also to detect in real
  time deviations in parameter values.
\end{itemize}

A notable trend in recent research on demand-responsive traffic
management systems is greater reliance on artificial intelligence (AI)
methods, prompted by an ever growing complexity of algorithms, models
and data. Let us cite some examples of this trend:
\cite{aut-liw04,aut-say98,aut-nii03} and CLAIRE \cite{aut-sce94}.

Overall, no matter how sophisticated these classical approaches,
\begin{itemize}
\item either their responsiveness is limited and they appear as tools
  both coercive and normative (imposing a traffic situation rather
  than responding to it),
\item or they are completely demand-responsive (CLAIRE or PRODYN for
  instance) and lack a global coordination. The TUC strategy might be
  viewed as a nice compromise.
\end{itemize}
All classical approaches require vast amounts of data collection and
processing, as well as huge processing power. Further, global
coordination notoriously requires data difficult to obtain or elaborate
such as dynamic origin-destination matrices or dynamic assignment data. Finally, the
systems described so far have a difficult time responding to
exceptional events, accidents, temporary building sites or other changes in the road network,
natural or industrial disasters, catastrophes, terrorist attacks etc.

Hence the usefulness of the decentralized and self-organized approach
advocated in this paper is its greater degree of flexibility, its independence of
a central traffic control center, and its greater robustness with respect to local perturbations
or failures. As shown in Sec.~\ref{trafcont} and summarized in Sec.~\ref{summa},
our autonomous adaptive control based on a traffic-responsive self-organization of traffic lights
leads to reasonable operations, including synchronization patterns such as green waves.
In particular, our principle of self-control is suited for
irregular (i.e.\ non-Manhattan type) road networks with counterflows, with main roads (arterials) and side roads,
with varying inflows, and with changing turning or assignment fractions. This distinguishes our
approach from simplified scenarios investigated elsewhere
\cite{JamsInCities,IsolatedCrossroads,RandomTrafficLights}. Another
interesting feature is that our approach considers not only ``pressures'' on the
traffic lights related to delay times. It also takes into account ``counter-pressures''
when subsequent road sections are full, i.e.
when green times cannot be effectively used.

\section{Modeling traffic flow in urban road networks}

In our model of urban road traffic, road networks are composed
of nodes (intersections, plazas, dead ends, or cross sections of the road), which
are connected by directed links $i$, representing homogeneous road sections without changes
in capacity.

\subsection{Traffic flow on network links}
\label{sec:Links}

\begin{figure}[htbp]
    \begin{center}
    \includegraphics[width=1\textwidth]{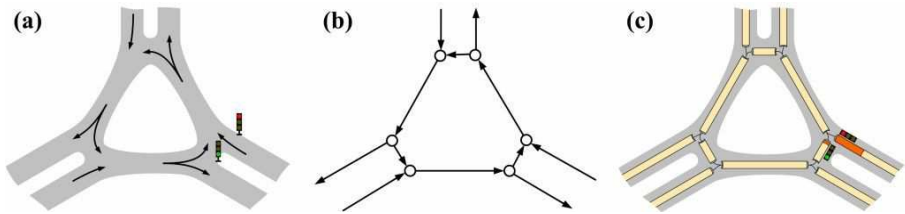}
    \end{center}
\caption{A road network (a) can be considered as a directed graph
(b). The directed links represent homogeneous road sections, while the nodes
correspond to junctions. (c) The road sections may or may not
be controlled by traffic lights.}
        \label{fig:Network}
\end{figure}

\subsubsection{Homogeneous road sections}

Our road sections $i$ are characterized by a constant number $I_i$ of lanes, over which
traffic is assumed to be equally distributed. Different lanes turning into different directions
may be treated as separate road sections, depending on the
respective design of the infrastructure. Road sections can have a very large length $L_i$,
which is in favor of numerical efficiency.
The dynamics within a link of the road network is described by
the section-based queueing-theoretical traffic model
by Helbing (2003b). It is directly related to the equation
of vehicle conservation \cite{LW} and briefly
introduced, here. The average velocity of vehicles on link $i$ around place $x$ at time $t$ is
denoted by $V_i(x,t)$, the spatial density per lane by $\rho_i(x,t)$, and the
flow per lane by $Q_i(x,t) = \rho_i(x,t) V_i(x,t)$. The flow is
approximated by a triangular flow-density relationship
\begin{equation}
 Q_i(x,t) = \left\{
\begin{array}{ll}
\rho_i(x,t) V_i^0 & \mbox{if } 1/\rho_i(x,t) > \left(1/\rho^{\rm jam} + T V_i^0\right) \, \\
\frac{1}{T}\left[ 1 - \rho_i(x,t)/\rho^{\rm jam} \right] & \mbox{otherwise (in congested traffic).}
\end{array}\right.
\label{flowdens}
\end{equation}
While the increasing line $\rho_i V_i^0$ describes free traffic moving with speed $V_i^0$,
the falling ``jam line'' describes congested traffic, in which the average vehicle distance
$1/\rho_i$ is given by an effective vehicle length $l^{\rm eff} = 1/\rho^{\rm jam}$
(= vehicle length plus  minimum front-bumper-to-back-bumper distance) plus a
safety distance $T V_i$ which grows linearly with the speed $V_i$. The proportionality factor
is the (safe) time gap $T$ kept in congested traffic. Therefore, our model is based on
only three intuitive parameters: the maximum
jam density $\rho^{\rm jam}$, the free velocity $V_i^0$ (speed limit)
on link $i$, and the time gap in congested traffic $T$. In our paper, we have chosen
$V_i^0=14 \mbox{ m/s} = 50$~km/h, $\rho^{\rm jam} = 150$~vehicles
per kilometer and lane, and $T = 1.8$~s.
\par
We should note that there are other macroscopic traffic models such as the
non-local, \underline{g}as-\underline{k}inetic-based \underline{t}raffic (GKT) model \cite{GKT},
which can describe the
aggregate dynamics of traffic flows more accurately than this model. The ``GKT model''
has even been successfully implemented to simulate traffic flows on all German
freeways, taking into account information by local detectors and floating car
data. However, the dynamics of urban traffic is dominated by the dynamics of the traffic
lights, which justifies simplifications in favor of numerical efficiency and
analytical treatment. The section-based traffic model covers the most
essential features of traffic flow in urban road networks, e.g.\ the transition between
free and congested traffic, the spreading and interaction of vehicle
queues, etc. Its particular strengths are its transparency,
numerical stability, and computational efficiency. Compared to microsimulation
models of urban traffic such as cellular automata models
\cite{Cremer,Esser,NagelSim}, the treatment of lane changes, intersections,
and turning operations is much easier, and analytical investigations are possible.

\subsubsection{Propagation of perturbations}

The particular simplicity of the section-based traffic model results from
its two constant characteristic velocities: While perturbations of free traffic
propagate together with the cars at the speed $V_i^0$, in congested traffic
perturbations travel upstream with the constant velocity
\begin{equation}
 c = - 1/(T \rho^{\rm jam}) \, ,
\end{equation}
which has the typical value  of $-3.7$~m/s or $-13.3$~km/h.

\begin{figure}[htbp]
    \begin{center}
        \includegraphics[width=0.7\textwidth]{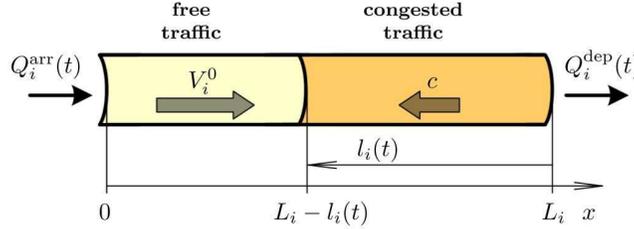}
    \end{center}
\vspace*{-4mm}
\caption{A road section $i$ of length $L_i$ with an area $l_i$ of
congested traffic at the downstream end (right). Due to the
constant propagation speeds $V_i^0$ and $c$ of
perturbations in free and congested traffic,
respectively (see big arrows), the internal dynamics can
be easily calculated based on the boundary flows
$Q_i^{\rm arr}(t)$ and $Q_i^{\rm dep}(t)$ only.}
        \label{fig:RoadSection}
\end{figure}

A favorable property of the section-based traffic model is that all relevant
quantities can be determined from the boundary flows, which makes the model
very efficient. For example, the
dynamics inside a road section $i$ can be easily derived from the arrival flow
$Q_i^{\rm arr}(t)$ and the departure flow $Q_i^{\rm dep}(t)$ per lane
with the two characteristic velocities $V_i^0$ and $c$,
see Fig.~\ref{fig:RoadSection}.
\par
The interior flow per lane is given by
\begin{equation}
 Q_i(x,t) = \left\{
\begin{array}{ll}
Q_i^{\rm arr}\left(t - \frac{x}{V_i^0}\right) & \mbox{if }
 x < L_i -l_i(t) \mbox{ (in free traffic)}\, , \\
Q_i^{\rm dep}\left(t - \frac{L_i-x}{|c|}\right) & \mbox{if } L_i - l_i(t) \le x \le L_i.
\end{array}\right.
\end{equation}
That is, the flow is determined by the downstream boundary
in the area of congested traffic of length $l_i(t) \ge 0$,
while it is given by the arrival flow in the area $x < L_i - l_i(t)$ of free traffic.
The density can be obtained via
\begin{equation}
 \rho_i(x,t) = \left\{
\begin{array}{ll}
 Q_i(x,t)/V_i^0 & \mbox{if } x < L_i -l_i(t) \mbox{ (in free traffic)}\, , \\
 {[1 - TQ_i(x,t)]} \rho^{\rm jam} & \mbox{if } L_i - l_i(t) \le x \le L_i.
\end{array} \right.
\end{equation}
The average velocity is calculated via the formula
$V_i(x,t) = Q_i(x,t)/\rho_i(x,t)$, if $\rho_i(x,t) > 0$.
\par
The temporal change of the number $N_i(t)$ of vehicles per lane on road section $i$ can be also
determined from the arrival and departure flows:
\begin{equation}
 \frac{dN_i}{dt} = Q_i^{\rm arr}(t) - Q_i^{\rm dep}(t) \, .
\label{num}
\end{equation}
The time-dependent change
of the congested area of length $l_i(t)$ will be discussed in the next paragraph.

\subsubsection{Movement of congestion fronts}
\begin{figure}[htbp]
    \begin{center}
        \includegraphics[width=0.7\textwidth]{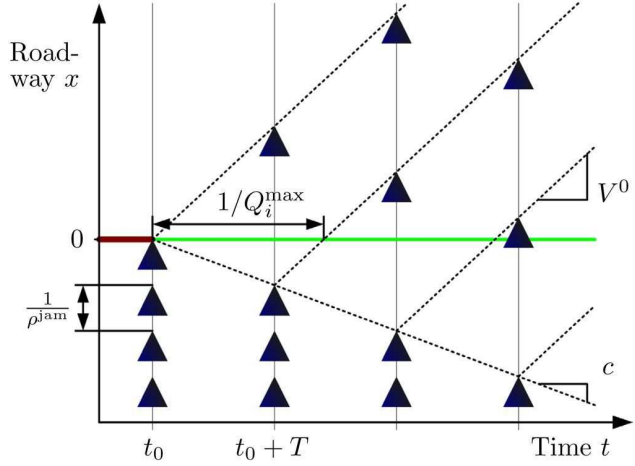}
    \end{center}
\caption{Illustration of queued vehicles (triangles in the lower left corner) and freely moving vehicles
after a traffic light turns green (triangles in the upper right part). The characteristic speeds $V^0$ and $c$
are indicated by diagonal lines.}
        \label{fig:corr}
\end{figure}

Since our road sections are homogeneous by definition, congestion
can only be triggered at their downstream ends. While the congested
area might eventually expand over the entire road
section, the downstream end remains at $x=L_i$. The upstream end
lies at $x=L_i - l_i(t)$, where jumps $\Delta \rho_i$ and $\Delta Q_i$
occur in the density and in the flow, respectively. In order to
ensure the conservation of vehicles, the condition $\Delta Q_i =
-\Delta \rho_i \cdot dl_i/dt$ must be fulfilled. Therefore, the border
line between free and congested traffic moves with the following velocity \cite{SectionBasedModel}:
\begin{equation}
    \frac{dl_i}{dt}
    \;=\;
    -\;\frac{Q_i^{\rm arr}\big(t-[L_i-l_i(t)]/V_i^0\big)
        - Q_i^{\rm dep}\big(t-l_i(t)/|c|\big)}
    {\rho_i^{\rm arr}\;\big(t-[L_i-l_i(t)]/V_i^0\big)
        - \rho_i^{\rm dep}\;\big(t-l_i(t)/|c|\big)}
    \;.
\label{queulen}
\end{equation}
Note that, within the congested area of length $l_i(t)$, one might find areas of quasi-free
traffic, where the vehicles reach the maximum free velocity $V_i^0$ and
the maximum flow $Q_i^{\rm max}$ per lane that is possible according
to the flow-density relationship (\ref{flowdens}):
\begin{equation}
    \label{eq:MaximumFlow}
    Q_i^{\rm max} =
    \left(T + \frac{1}{V_i^0 \rho^{\rm jam}}\right)^{-1} \, .
\end{equation}
This value corresponds to vehicles accelerating out of a traffic jam every
$T=1.8$ seconds. Nevertheless, the value $1/T$ is not completely reached,
as each subsequent vehicle has to drive an additional distance
$l^{\rm eff} = 1/\rho^{\rm jam}$ in order to reach the respective measurement
cross section. This requires an additional time interval of $l^{\rm eff}/V_i^0$
as in the formula above (see Fig.~\ref{fig:corr}).
\par
Let us shortly discuss two special cases of formula (\ref{queulen}): If the departure flow
is stopped due to a red traffic light, we obtain the simplified relationship
\begin{equation}
 \frac{dl_i}{dt} = \left[ \frac{\rho^{\rm jam}}{Q_i^{\rm arr}\left(t-[L_i-l_i(t)]/V_i^0\right)}
 - \frac{1}{V_i^0} \right]^{-1}
\approx \frac{Q_i^{\rm arr}\left(t-[L_i-l_i(t)]/V_i^0\right)}{\rho^{\rm jam}}  \, .
\label{spe}
\end{equation}
If the traffic light turns green at time $t'_0$, the end of the traffic jam still propagates upstream at the speed
(\ref{spe}) with new arriving vehicles. However, at the same time,
an area of quasi-free traffic with maximum flow
$Q_i^{\rm max}$ propagates upstream with velocity $c$ from the downstream boundary.
Therefore, the effective length $l_i^{\rm eff}(t)$ of the vehicle queue is
\begin{equation}
 l_i^{\rm eff}(t) = l_i(t) - |c| (t-t'_0) \, .
\end{equation}
If this effective queue has been fully resolved at time $t^*$, i.e.\ $l_i^{\rm eff}(t^*) = 0$,
it takes an additional time $l_i(t^*)/V_i^0$ until the last vehicle of that queue has left
the road section $i$. Therefore, we reach $l_i(t)=0$ and, thereby,
free traffic on the whole road section $i$, at time $t^*+l_i(t^*)/V_i^0$. Before this
point in time, vehicles that have moved out of the queue
may still be trapped again by a red traffic light at the end of road section $i$.

\subsubsection{Travel time}
Let the travel time $T_i(t)$ be the time a vehicle needs to
pass through the road section $i$ when entering it at time $t$. Then,
the actual number $N_i(t)$ of vehicles inside the road section is
given by
\begin{equation}
    \label{eq:NumberOfVehicles_Cumulated}
    N_i(t) = \!\!
    \int\limits_{t}^{t+T_i(t)} \!\! dt' \; Q_i^{\rm dep}(t') \, .
\end{equation}
This formula implies the
following delay-differential equation describing how the travel time $T_i$
depends on the boundary flows \cite{SectionBasedModel}:
\begin{equation}
    \label{eq:TravelTime}
    \frac{dT_i}{dt} =
    \frac{Q_i^{\rm arr}(t)}{Q_i^{\rm dep}\big(t+T_i(t)\big)} - 1 \, .
\end{equation}
According to this, the travel time can be predicted based on the anticipated
departure flow, e.g.\ when a certain traffic light control is assumed
(see Secs.~\ref{DElay} and \ref{BAsic}).

\subsubsection{Delay time}\label{DElay}
Since the travel time would exactly be $L_i/V_i^0$ without
congestion, any deviation from that can be understood as the time
a vehicle has been delayed due to congestion. Therefore, we may
introduce the delay time
\begin{equation}
T_i^{\rm del}(t) = T_i - \frac{L_i}{V_i^0} \, .
\end{equation}
Since $L_i/V_i^0$ is time-independent, the right hand side of equation
(\ref{eq:TravelTime}) applies to $d T_i^{\rm del}/dt$ as well.

Consider a road section with a constant arrival flow
$Q_i^{\rm arr}(t)$ and a departure flow $Q_i^{\rm dep}(t)
= \gamma_i(t) Q_i^{\rm max}$ being controlled by a traffic light.
As the buffer size is given by the maximum number $L_i \rho^{\rm jam}$
of vehicles per lane on road section $i$, from Eq.~(\ref{num})
we can derive
\begin{eqnarray}
    \frac{1}{t} \int\limits_0^t dt' \; Q_i^{\rm arr} (t')
    &\leq & \frac{L_i \rho^{\rm jam}}{t} + \frac{1}{t}
    \int\limits_0^t dt' \; Q_i^{\rm dep}(t') \nonumber \\
    &\leq & \frac{L_i \rho^{\rm jam}}{t} + \frac{Q_i^{\rm max}}{t} \int\limits_0^t dt' \;
    \gamma_i(t') \nonumber \\
    &= & \frac{L_i \rho^{\rm jam}}{t} + u_i Q_i^{\rm max}
\end{eqnarray}
with the average green time fraction
\begin{equation}
  u_i = \frac{1}{t} \int\limits_0^t dt' \; \gamma_i(t') \, .
\end{equation}
For $t\rightarrow \infty$ we can see that the average arrival rate
per lane on road section $i$ should not exceed the maximum flow times the green time fraction
$u_i$. Otherwise, we will have a growing queue, until the maximum storage capacity
$I_iL_i\rho^{\rm jam}$ for vehicles on road section $i$ has been reached.
\par
\begin{figure}[htbp]
    \begin{center}
        \includegraphics[width=1\textwidth]{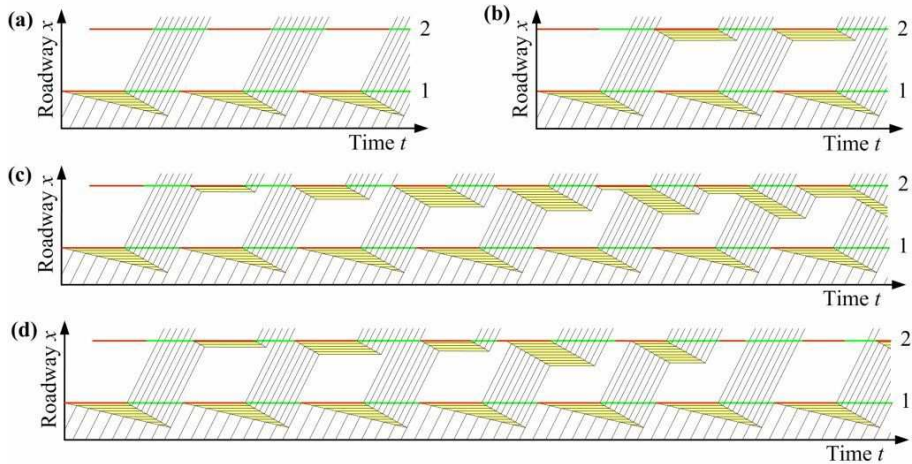}
    \end{center}
\caption{Trajectories of freely moving vehicles (diagonal lines) and queued
vehicles (horizontal lines) in dependence of the traffic light control at two
subsequent intersections 1 and 2. In all four displayed scenarios, vehicles arrive with identical time headways
(i.e.\ constant arrival rate) at traffic light 1, which operates periodically.
Traffic light 2 is operated in different modes:
(a) The frequency and time offset are adapted to the first traffic light, as required by a green
wave. (b) The frequency is the same as for the first traffic light, but has a non-optimal time offset.
(c) The frequency (and cycle time) differs from the one of the first traffic light. (d) The green time varies
stochastically, but the average green time fraction is the same. When the frequencies are the same, but the
time offset is not properly adjusted, a certain fraction of vehicles is stopped, see (b). If the frequencies are different, it is most
likely that vehicles will be stopped by a red light, potentially even for several times, see (c). In such cases,
a stochastic variation of green time periods can be favorable, see (d).}
        \label{fig:sync}
\end{figure}

The throughput is reduced if a downstream road section $j$
is sometimes fully congested, as this limits the departure flow. Moreover, the delay time
can temporarily increase, if the arrival of vehicles at the upstream boundary
of road section $i$ is not synchronized with the green phase of the traffic light at the downstream end.
Such a synchronization of arrivals in $i$ with the desired departure times is hard to reach
in an irregular road network. As a consequence, vehicles tend to queue up at a red light before
they can leave a road section $i$ (see Fig.~\ref{fig:sync}). Note, however, that a green light reaches
maximum efficiency when it serves vehicles which have queued up before.
\par
Let us now study the case where the waiting queues cannot be cleared completely
within one green phase. How long is a vehicle
delayed, if it joins a queue of length $l_i(t_0)$ at time $t_0$? The totally required
green time needed until the vehicle can leave the road section $i$ is given by
\begin{equation}
 T_i^{\rm req}(t_0) = \frac{l_i(t_0) \rho^{\rm jam}}{Q_i^{\rm max}} \, ,
\end{equation}
since $l_i(t_0) \rho^{\rm jam}$ is the number of vehicles per lane to be served and
$Q_i^{\rm max}$ the service rate. Let us now estimate the overall time passed
until the downstream boundary of road section $i$ is reached. It is given by the formula
\begin{equation}
 T_i^{\rm pass}(t_0) = T_i^{\rm req}(t_0) + \mbox{overall red and yellow times in between.}
\end{equation}
\par
The time delay of vehicle $i$ by queuing, red and yellow times is the overall time passed minus
the travel time $l_i(t_0)/V_i^0$ in free traffic:
\begin{eqnarray}
 T_i^{\rm del}(t_0) &=& T_i^{\rm pass} - \frac{l_i(t_0)}{V_i^0} \\
 &=& l_i(t_0) \left( \frac{\rho^{\rm jam}}{Q_i^{\rm max}} - \frac{1}{V_i^0} \right)
+ \mbox{overall red and yellow times.} \nonumber
\end{eqnarray}
Generally, this formula is difficult to express, as its result depends sensitively on
the respective red and green phases. However, the formula for the average delay time
becomes quite simple. Just remember that the average green time fraction is $u_i$ and
the average fraction of red and yellow times must be $1-u_i$. Therefore, the average
delay $\overline{T_i^{\rm del}}$ as a function of the average queue length
$\overline{l_i}$ and the green time fraction $u_i$ is estimated by the formula
\begin{eqnarray}
\overline{T_i^{\rm del}} &\approx &  \overline{l_i} \left( \frac{\rho^{\rm jam}}{Q_i^{\rm max}} - \frac{1}{V_i^0} \right)
+ \frac{1-u_i}{u_i} \times \mbox{totally required green time } T_i^{\rm req} \nonumber \\
&=& \overline{l_i} \left( \frac{\rho^{\rm jam}}{u_i Q_i^{\rm max}} - \frac{1}{V_i^0} \right) \, .
\end{eqnarray}
According to this, the average delay time $\overline{T_i^{\rm del}}$ is proportional to the average
queue length $\overline{l_i}$, but a large green time fraction $u_i$ is helpful.
Note that the formulas of this section are not only applicable to situations with
fixed cycle times and signal programs. They are also applicable to situations where
the red and green phases are varying.

\subsubsection{Potential flows and traffic states}
\label{sec:PotentialFlows} The in- and outflow of a road section is not only
limited by capacity constraints such as $Q_i^{\rm max}$, but also by
the actual state of traffic. We will, therefore, denote the potential arrival
and departure flows per lane by $Q_i^{\rm arr, pot}(t)$ and $Q_i^{\rm dep, pot}(t)$, respectively.
Congestion is triggered if $Q_i^{\rm dep}(t) > Q_i^{\rm dep,
pot}(t)$, and resolved if $l_i(t)=0$. In the case where the road
section is entirely congested, i.e.\ $l_i(t) = L_i$, this state remains
until $Q_i^{\rm arr}(t) < Q_i^{\rm arr, pot}(t)$. The potential
flows are determined as follows: As long as there is no
congestion, the potential departure flow is given by the former
arrival flow $Q_i^{\rm arr}(t-L_i/V_i^0)$.
When the downstream end of road section $i$ is congested, vehicles
are queued up and can depart with the maximum possible flow $Q_i^{\rm max}$.
Altogether, we have
\begin{equation}
 Q_i^{\rm dep, pot}(t) = \left\{
\begin{array}{ll}
Q_i^{\rm arr}(t-L_i/V_i^0) & \mbox{if } l_i(t) = 0\, , \\
Q_i^{\rm max} & \mbox{if } l_i(t) > 0 \, .
\end{array}\right.
\end{equation}
At the upstream end, the maximum possible flow $Q_i^{\rm max}$ can enter road section $i$ as long as it
is not entirely congested. Otherwise, the arrival flow is
limited by the former departure flow $Q_i^{\rm dep}(t-L_i/|c|)$. This implies
\begin{equation}
Q_i^{\rm arr, pot}(t) = \left\{
\begin{array}{ll}
Q_i^{\rm max} & \mbox{if } l_i(t) < L_i \, , \\
Q_i^{\rm dep}(t-L_i/|c|) & \mbox{if } l_i(t) = L_i \, .
\end{array}\right.
\end{equation}
In cases, where the outflow
of the road section is to be controlled by a traffic light,
the potential departure flow $Q_i^{\rm dep, pot}(t)$ must be multiplied with a
prefactor $\gamma_i(t)$. A green light corresponds to
$\gamma_i(t) = 1$, a red light to $\gamma_i(t) = 0$. Note that it is also
possible to vary $\gamma_i(t)$ gradually to account for drivers
passing the signal during yellow phases.

\subsection{Traffic flows through network nodes}
\label{sec:Nodes}

\begin{figure}[htbp]
\begin{center}
    \includegraphics[width=1\textwidth]{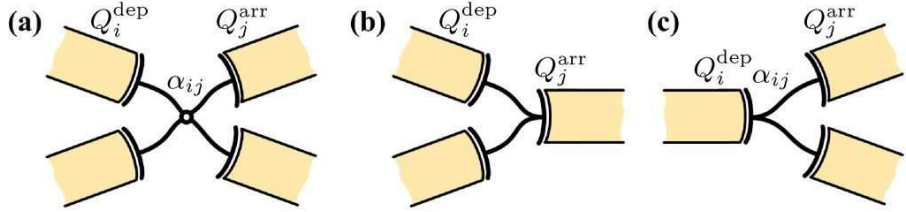}
\end{center}
\caption{(a) A node of the road network distributes the vehicular
flows between the road sections that are connected to it. It makes sense to distinguish
two special cases: (b) merges into a single road section and (c) diverges from one road section into
several others.}
        \label{fig:node}
\end{figure}

A node of the road network connects one or several incoming road sections
$i$ with one or several outgoing road sections $j$, see figure
\ref{fig:node}(a). It may represent a junction or a link of two subsequent homogeneous
road sections $i$ and $i+1$ with different speed
limits $V_i^0$, $V_{i+1}^0$ or numbers $I_i$, $I_{i+1}$ of lanes. Since nodes are assumed to have
no storage capacity, the total in- and outflow have to be the same (Kirchhoff's law):
\begin{equation}
    \label{eq:Kirchhoff}
    \underbrace{\Big.\sum\nolimits_i Q_i^{\rm dep}(t)}_{\rm inflow}
    = \underbrace{\Big.\sum\nolimits_j Q_j^{\rm arr}(t)}_{\rm outflow} \, .
\end{equation}
Furthermore, the flows have to be non-negative and must not exceed
the potential flows specified in Sec.~\ref{sec:PotentialFlows}.
\begin{equation}
    \label{eq:NodeLimitation}
        0 \leq Q_i^{\rm dep}(t) \le Q_i^{\rm dep,pot}(t) \, , \qquad
        0 \leq Q_j^{\rm arr}(t) \le Q_j^{\rm arr, pot}(t) \, .
\end{equation}
The fraction of the inflow $Q_i^{\rm dep}$ that diverges from road
section $i$ to road section $j$ is denoted by
$\alpha_{ij}(t)$. Due to normalization we have
\begin{equation}
         \sum_j \alpha_{ij}(t) = 1 \, .
\end{equation}
The turning or assignment coefficients $\alpha_{ij}$ may depend on the
driver destinations $d$ as well as on the actual traffic
situation, see Daganzo (1995) and Sec.~\ref{sec:TrafficAssignment}.
Finally, note that the arrival flow $Q_j^{\rm arr}(t)$ is composed of
all turning flows $Q_i^{\rm dep}(t) \alpha_{ij}(t)$ entering road section $j$:
\begin{equation}
\label{eq:DivergeFraction}
        Q_j^{\rm arr}(t) = \sum_i Q_i^{\rm dep}(t) \alpha_{ij}(t)  \, .
\end{equation}
For a more detailed treatment of network nodes see Lebacque (2005).

\subsubsection{Merges}
In the case where traffic flows from several incoming road sections $i$
merge into one outgoing road section $j$, as shown in
Fig.~\ref{fig:node}(b), two cases can be distinguished: As
long as the subsequent road section $j$ has sufficient capacity to
admit the potential flows of all incoming road sections $i$, i.e.
$Q_j^{\rm arr, pot}(t) \geq \sum_i Q_i^{\rm dep,pot}(t)$, the flow
through the node is given by the upstream traffic conditions in the road sections $i$.
Otherwise, some of the upstream departure flows $Q_i^{\rm dep}(t)$ have to be
restricted. But which ones? According to practical experience, small traffic flows
$Q_i^{\rm dep}(t)$ can almost always squeeze in, while flows from equivalent roads
tend to share the capacity $Q_j^{\rm arr, pot}$ equally.
Note that in scenarios with main roads having a right of way, the corresponding flow
is to be served first. The remaining capacity is subsequently distributed
among the side roads.

\subsubsection{Diverges}
Figure \ref{fig:node}(c) shows the case where traffic diverges
from one road section into several others. This is, for example, the case
when a road splits up into lanes for turning left, continuing straight ahead, or turning right.
For diverges, the throughput is determined by a cascaded minimum-function:
\begin{equation}
    \label{eq:DivergingNode}
    Q_i^{\rm dep}(t) = \min \left\{
    Q_i^{\rm dep, pot}(t), \;\;
    \min_{j} \frac{Q_{j}^{\rm arr,pot}(t)}{\alpha_{i{j}}(t)}
    \right\} \, .
\end{equation}
The first term on the right-hand side is obvious, as any restriction of the potential
departure flow $Q_i^{\rm dep, pot}(t)$ of road section $i$
limits the flows to all outgoing road sections $j$. The second term on the right-hand
side follows from the fact that the fraction $\alpha_{ij}$ of the departure flow
$Q_i^{\rm dep}(t)$ to any subsequent road section $j$ is limited by its potential
arrival flow $Q_j^{\rm arr, pot}(t)$, i.e.
\begin{equation}
 Q_i^{\rm dep}(t) \alpha_{ij} \le Q_j^{\rm arr, pot}(t) \quad \forall j \, .
\end{equation}
In the special case of a node connecting only two subsequent road sections
$i$ and $j=i+1$, we have $\alpha_{ij}=1$ and the throughput is just limited
by the minimum of both potential flows:
\begin{equation}
    \label{eq:SimpleNode}
    Q_i^{\rm dep}(t)
  =  \min \Big\{ Q_i^{\rm dep, pot}(t),\;\; Q_{i+1}^{\rm arr, pot}(t) \Big\}  = Q_{i+1}^{\rm arr}(t) \, .
\end{equation}
The last equality follows from Eq.~(\ref{eq:DivergeFraction}).

\section{Traffic assignment} \label{sec:TrafficAssignment}

The simplest way to model turning at intersections is by turning coefficients $\alpha_{ij}(t)$, which
assume that a certain fraction $\alpha_{ij}(t)$ of the departure flow $Q_i^{\rm dep}(t)$ turns into
road section $j$. In many theoretical studies, the coefficients $\alpha_{ij}$ are kept constant. However, it is
well-known that the turning fractions vary in the course of the day, which is often taken into account
by using historical, time-dependent turning coefficients $\alpha_{ij}(t)$ from a database
\cite{SchreckenbergData}. Moreover,
even if the same origin-destination flows would repeat each week, delays due to perturbations
in the traffic flow (e.g.\ due to an accident) would cause different time-dependent turning fractions.
Therefore, a better treatment is based on dynamic traffic assignment.
\par
In order to integrate dynamic traffic assignment in our model, let us denote the destination node
of vehicles by $d$. Moreover, let $N_{id}(t)$ represent the number of driver-vehicle units on
the directed link $i$, which finally want to arrive at $d$. This implies
\begin{equation}
 N_{i}(t) = \sum_d N_{id}(t) \, .
\end{equation}
The quantity $Q_{id}^{\rm arr}(t)$ shall denote the flow of vehicles with destination $d$
entering the link $i$, and $Q_{id}^{\rm dep}(t)$ the flow of vehicles leaving it. We have
\begin{equation}
 Q_{i}^{\rm arr}(t) = \sum_d Q_{id}^{\rm arr}(t) \quad \mbox{and} \quad
 Q_{i}^{\rm dep}(t) = \sum_d Q_{id}^{\rm dep}(t) \, .
\end{equation}
Finally, let $\overline{j}$ be
the starting node of link $j$ and $\underline{j} = k$ its ending node.
Moreover, let $T_{\overline{j}k}(t)$ be the travel time on link $j$ and
$\widehat{T}_{kd}(t)$ the minimum travel time between two nodes $k$ and $d$
(as can, for example, be determined by the Dijkstra algorithm). Then,
the minimum travel time to note $d$ via link $j$ (i.e.\ node $k$)
is given by $T_{\overline{j}k}(t) + \widehat{T}_{kd}(t)$, and
the minimum travel time $\widehat{T}_{\overline{j}d}(t)$ from node $\overline{j}$
to destination $d$ at time $t$ is determined via
\begin{equation}
  \widehat{T}_{\overline{j}d}(t) = \min_k [ T_{\overline{j}k}(t) + \widehat{T}_{kd}(t)] \, ,
\label{dij}
\end{equation}
where the minimum function extends over all successors $k$ of node $\overline{j}$. Instead of this,
we may use the following approximate relationship:
\begin{equation}
  \widehat{T}_{\overline{j}d}(t) = \min_k [ T_{\overline{j}k}(t) + \widehat{T}_{kd}(t- \Delta t)] \, .
\label{approx}
\end{equation}
The advantage of (\ref{approx}) over (\ref{dij}) is that the information
about travel times gradually propagates to the present location
of the car (namely by one link each time step $\Delta t$).
A delayed evaluation of Dijkstra's shortest path algorithm saves computer time and
models this information flow, the speed of which is controlled by $\Delta t$. Another
advantage is the determination of travel times based on a local algorithm.
\par
Based on this travel time information, we may
distribute the departure flows $Q_i^{d,{\rm dep}}(t)$
over neighboring links according to a multinomial logit model \cite{MNL}. Accordingly,
we specify the turning probabilities of cars with destination $d$ at node $\overline{j}= \underline{i}$ as
\begin{equation}
 p_{\overline{j}k}^d(t) = \frac{\mbox{exp} \{-\beta  [T_{\overline{j}k}(t) + \widehat{T}_{kd}(t-\Delta t)]/\widehat{T}_{\overline{j}d}^0 \}}
{\sum_{k'} \mbox{exp} \{-\beta  [T_{\overline{j}k'}(t) + \widehat{T}_{k'd}(t-\Delta t)]/\widehat{T}_{\overline{j}d}^0\}} \, ,
\end{equation}
where $\widehat{T}_{\overline{j}d}^0$ is the minimum travel time from $\overline{j}$ to $d$
during free traffic (at three o'clock during the night).
The coefficient $\beta$ describes the sensitivity with respect to changes in the relative travel time
and is also a measure for the reliability of travel time estimates.
Finally, the time-dependent assignment coefficients can be calculated as
\begin{equation}
 \alpha_{ij}(t) = \sum_d \frac{Q_{id}^{\rm dep}(t)}{Q_i^{\rm dep}(t)} \, p_{\underline{i}\underline{j}}^d(t) \, ,
\end{equation}
where $\underline{i} = \overline{j}$ and $\underline{j} = k$.
This assumes individual route choice decisions without central coordination, i.e.\ selfish routing.
\par
We must still decide how to determine travel times.
On the one hand, one may use the expected travel times $T_{\overline{j}k}(t) = T_{\overline{j}\underline{j}}(t)
= T_j(t)$ according to Eq.~(\ref{eq:TravelTime}) (or, as a second best alternative,
the instantanous link travel times). On the other hand, one may use travel time information
$T_{\overline{j}k}^*(t)$ of comparable days from a database \cite{SchreckenbergData}.
While for close links, the expected travel time may be a good (and the instantaneous travel time
a reasonable) estimate of the actual travel time, it becomes less reliable
the more remote the respective link is. For remote links, a travel time estimate based on measurements
of similar previous days may be more reliable. Therefore,
we propose to use a weighted mean value generalizing formula (\ref{approx}):
\begin{equation}
  \widehat{T}_{\overline{j}d}(t) = \min_k [ T_{\overline{j}k}(t)
+ \mbox{e}^{-\lambda\,T_{\overline{j}k}(t)} \widehat{T}_{kd}(t- \Delta t)
 + (1 - \mbox{e}^{-\lambda\,T_{\overline{j}k}(t)}) T_{kd}^*(t)] \, .
\end{equation}
In this formula, the travel time $T_{kd}^*(t)$ from node $k$ to $d$ is taken from a database,
the weights are exponentially decaying with increasing travel times, and
$\lambda > 0$ is a suitably chosen calibration parameter.
\par
Right now it is not clear what happens if traffic lights adapt to the traffic situation and
drivers try to adjust to the traffic lights at the same time. Driver adaptation is a reasonable
strategy for signal plans that are fixed or determined by the time of the day. However, it
may perturb attempts to optimize traffic by self-organized control. Therefore, the study
of route choice behavior in the context of adaptive traffic light control requires careful study.
A method to stabilize the system dynamics, if needed, would be road pricing (see Sec.~\ref{pricing}).

\section{Self-organized traffic light control} \label{trafcont}

\subsection{Why traffic lights?}\label{why}

For the illustration of the advantages of oscillatory traffic control, let us
assume a conventional four-armed intersection with identical capacities $Q_i^{\rm max}
= Q^{\rm max}$.
The arrival time of vehicles shall be stochastic. Vehicles are assumed
to obstruct the intersection area (i.e.\ the node) for a time period of $1/Q^{\rm max}$ in case of
compatible flow directions. For incompatible, e.g.\ crossing flows,
the blockage time shall be $\tau = sT$ with $s>1$. The maximum average
throughput ${Q^{\rm cap}}$ of the intersection is, therefore, bounded by the following inequality:
\begin{equation}
 \frac{1}{T} > Q^{\rm max} \ge {Q^{\rm cap}} \ge \frac{1}{\tau} = \frac{1}{sT} \, .
\end{equation}
The exact value of ${Q^{\rm cap}}$ depends on the fractions of compatible and
incompatible flows. For compatible flows only, we have ${Q^{\rm cap}}
= Q^{\rm max}$. If the vehicle flows were always incompatible, one would have ${Q^{\rm cap}} = 1/\tau = 1/(sT)$.
\par
Let us now cluster vehicles into platoons of $n$ vehicles by the use of suitable adaptive traffic
lights. Moreover, let the green phases last for the time periods $\Delta \tau_i$. Between the
green periods, we will need yellow lights for a time period of $\tau$ to prevent accidents. An
estimate of the capacity ${Q^{\rm cap}}$ of the signalized intersection is then
\begin{equation}
 {Q^{\rm cap}} = \frac{\sum_{i=1}^{k} Q_i^{\rm max} \Delta \tau_i}
{\sum_{i=1}^{k} (\Delta \tau_i + \tau)}
= Q^{\rm max} \frac{\sum_{i=1}^{k}  \Delta \tau_i} {{T^{\rm cyc}}} \, ,
\label{capa}
\end{equation}
where ${T^{\rm cyc}} = {k}\tau + \sum_i \Delta \tau_i$ is the average cycle time.
Of course, there are different possible schemes to control the intersection, but we can show
that for $n$-vehicle platoons with $\Delta \tau_i = n/Q^{\rm max}$,
the capacity of the signalized intersection is
\begin{equation}
 {Q_{(n)}^{\rm cap}} = \frac{kn/Q^{\rm max}}{kn/Q^{\rm max} + ksT}
 = \left(\frac{1}{Q^{\rm max}} + \frac{sT}{n}\right)^{-1} \, .
\end{equation}
This is greater than the
capacity $1/(sT)$ of an uncontrolled intersection
with incompatible flows, if
\begin{equation}
 sT \left( 1 - \frac{1}{n} \right) > \frac{1}{Q^{\rm max}} \ge T \, ,
\end{equation}
i.e.\ if $s$ or $n$ are large enough. In other words: Forming vehicle platoons (clusters)
by oscillatory traffic lights can increase the intersection capacity. This, however, requires
that the green times are fully used. Otherwise, at small arrival rates, traffic lights would potentially
delay vehicles.
\par
Despite of the simplifications made in the above considerations, the following
conclusions are quite general: It is most efficient if vehicles can
pass the intersection immediately one by one, if the arrival rates are small. Above a
certain threshold, however, it is more efficient to form vehicle platoons by means of traffic lights.
This is certainly the case, if the sum of arrival flows exceeds the
capacity of an unsignalized intersection with incompatible flows.
According to formula (\ref{capa}), the capacity of a signalized intersection can be increased by
increasing the green time fractions $\Delta \tau_i/{T^{\rm cyc}}$. This can
be done by increasing the cycle time $T^{\rm cyc}$ in cases of high arrival flows
${Q_i^{\rm arr}}$. Thereby, the relative blockage time by yellow lights is reduced.

\subsection{Self-induced oscillations}

In pedestrian counterflows at bottlenecks, one can often observe
oscillatory changes of the passing direction, as if the pedestrian flows were
controlled by a traffic light. Inspired by this, we have suggested to generalize this principle
to the self-organized control of intersecting vehicle flows [see the newspaper article
by Stirn (2003)]. This
idea was described in 2003 in the DFG proposal He 2789/5-1 entitled ``Self-organized
traffic signal control based on synchronization phenomena in driven many-particle systems
and supply networks''.
The control concept elaborated in the meantime has been submitted for a patent.
For visualizations of some traffic scenarios
see the videos available at {\tt www.trafficforum.org/trafficlights/}.
\par\begin{figure}[htbp]
    \begin{center}
        \includegraphics[width=1\textwidth]{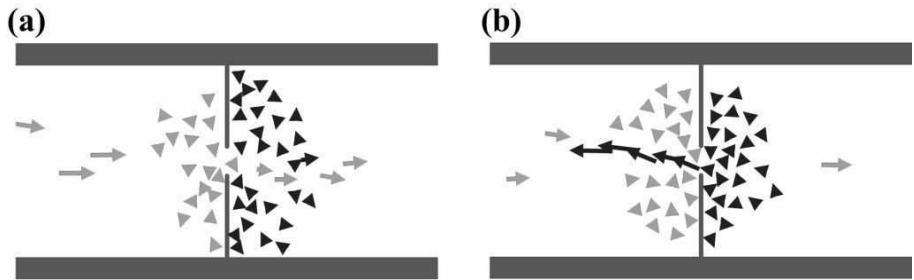}
    \end{center}
\caption{Alternating pedestrian flows at a bottleneck. These
oscillations are self-organized and occur due to a pressure difference between the
waiting crowd on one side and the crowd on the other side passing the bottleneck
[after Helbing and Moln\'{a}r (1995), Helbing (1997)].}
    \label{fig:Pedestrians}
\end{figure}
Oscillations are a organization pattern of conflicting flows which allows to optimize the overall throughput
under certain conditions (see Sec.~\ref{why}). In pedestrian flows (see Fig.~\ref{fig:Pedestrians}),
the mechanism behind the self-induced oscillations is as follows:
Pressure builds up on that side of the bottleneck where
more and more pedestrians have to wait, while it is reduced on the side
where pedestrians can move ahead and pass the bottleneck. If the pressure on
one side exceeds the pressure on the other side by a certain amount, the
passing direction is changed.
\par
Transferring this self-organization principle to urban vehicle
traffic, we define red and green phases in a way that considers ``pressures'' on the traffic light
by road sections waiting to be served and ``counter-pressures'' from the subsequent road sections
depending on the degree of congestion on them. Generally speaking, these pressures depend on
delay times, queue lengths, or potentially other quantities as well. The proposed
control principle is self-organized, autonomous,
and adaptive to the respective local traffic situation, as will be shown below.

\subsection{Basic switching rules for traffic lights} \label{BAsic}

Our switching rules for traffic lights will have to solve the following control problems:
\begin{itemize}
\item The number of vehicles on a road section served by a green time period should
be proportional to the average arrival flows $\overline{Q_i^{\rm arr}}$, at least if these are small.
\item In order to avoid time losses due to yellow lights, switching of traffic lights
should be minimized under saturated traffic conditions.
However, single vehicles and small queues need to be served as well
after some maximum cycle time $T^{\rm max}$.
\item Despite of the desire to maintain green lights as long as possible, signal control
should be able to react to changing traffic conditions in a flexible way. Unfortunately,
the change of traffic conditions depends on traffic light control itself, so that a reliable forecast
is only possible over short time periods.
\item Under suitable conditions, traffic lights should synchronize themselves to establish green waves.
\end{itemize}
The synchronization of traffic lights is not
only a matter of the adjustment of green and red time periods,
i.e.\ of the frequency of control cycles: The adaptation of the time offset is
also crucial for the establishment of green waves. While the adaptation problem is easily
solvable for Manhattan-like road networks, the situation for irregular road networks is much
more complex. Green waves may, in fact, cause major obstructions of crossing flows. Therefore,
it is a great difficulty to find suitable rules which flows to prioritize.
While addressing these points in the next paragraphs, we will develop a suitable control
approach step by step. The resulting control principles may be also used to resolve conflicts between
competing flows in other complex systems like production networks
[Helbing (2003a, 2004, 2005), Helbing {\em et al.} (2004)],
see Sec.~\ref{product}.
\par
The philosophy of our traffic light control
is the minimization of the cumulative or average travel time and, therefore,
of the cumulative delay time.
Minimizing the overall delay time means to serve as many vehicles by the traffic lights as possible,
i.e.\ to maximize the average departure rate (the average throughput).
Let us explain this principle in more detail:
If the traffic light is red or yellow, we have $\gamma_i(t) = 0$ and the overall
departure rate is $I_iQ_i^{\rm dep}(t) = 0$.
Otherwise, if the traffic light is green ($\gamma_i(t) = 1$), we find
\begin{equation}
I_i Q_i^{\rm dep}(t) = \left\{
\begin{array}{ll}
I_i Q_i^{\rm arr}(t-L_i/V_i^0) & \mbox{if } l_i(t) = 0 \, , \\
\min_j [I_jQ_j^{\rm dep}(t-L_i/|c|)/\alpha_{ij}] & \mbox{if } l_j(t) = L_j \, , \\
I_i Q_i^{\rm max} & \mbox{otherwise.}
\end{array}\right.
\label{conds}
\end{equation}
A green light should be provided for the road section whose
vehicle flow during a certain future time period is expected to be highest, taking into account
any yellow-light related time losses. This principle tends to serve the road with the largest
outflow, i.e.\ the largest number $I_i$ of lanes (see the third condition). However, it matters
how long the maximum flow can be maintained, i.e.\ how large the number
number $I_il_i\rho^{\rm jam}$ of queued vehicles is.  Moreover,
vehicles in road section $i$ will be hardly able to depart (see the second condition),
if one of the subsequent road sections
$j$ is completely congested by the expected number
$I_i Q_i^{\rm max} \alpha_{ij} (t-t'_0)$ of vehicles arriving between time $t'_0$ and $t$.
That is, a green light starting at time $t'_0$ would usually end
when the condition
\begin{equation}
I_i Q_i^{\rm max} (t-t'_0) \alpha_{ij} = I_j [L_j -l_j(t'_0)]\rho^{\rm jam}
\end{equation}
is valid for the first time.
Freely moving vehicles (see the first conditions) will have an impact comparable to the
reduction of a queue (third condition) only, if
\begin{equation}
 \frac{1}{t-t_0} \int\limits_{t_0}^t dt' \; Q_i^{\rm arr}(t-L_i/V_i^0)
 = \frac{N_i^{\rm arr}(t-L_i/V_i^0) - N_i^{\rm arr}(t_0-L_i/V_i^0)}{t-t_0}
\end{equation}
is of the order $Q_i^{\rm max}$, where
\begin{equation}
 N_i^{\rm arr}(t-L_i/V_i^0) = \int\limits_{0}^t dt' \; Q_i^{\rm arr}(t'-L_i/V_i^0) \, .
\end{equation}
\par
Summarizing this, the expected number $\Delta N_i^{\rm exp}$ of vehicles
served before interruption by a red light at time $t_1$ can be often
estimated by the cascaded minimum function
\begin{eqnarray}
 \Delta N_i^{\rm exp} &=& I_i Q_i^{\rm max} \, (t_1-t'_0) \nonumber \\
 &=&  \rho^{\rm jam} \min\bigg[\!\underbrace{I_il_i(t'_0)}_{\rm pressure} \, ,
\underbrace{\min_j \bigg(\frac{I_j[L_j-l_j(t'_0)]}{\alpha_{ij}}
\bigg) }_{\rm counter-pressure} \! \bigg] \, ,
\label{swrule}
\end{eqnarray}
where $t_1-t'_0$ denotes the expected green time.
However, generalizations of this formula are needed for the treatment of low traffic
(see Sec.~\ref{single}) and green waves (see Sec.~\ref{green}).
\par
As our control philosophy
requires to reduce queues as fast as possible,
the decision to serve a certain road section $i$ should be
based on the greatest value of $\sum \Delta N_i^{\rm exp}/(t_1-t'_0)$, where the
sum extends over all flows compatible with $Q_i^{\rm dep}$. If a switching time
$\tau$ is necessary, the relevant formula is $ \sum \Delta N_i^{\rm exp}/(t_1-t'_0+\tau)$,
instead. The switching decision should be regularly revised (e.g.\ every time period $\tau$),
as the traffic situation may change.
\par
Note that formula (\ref{swrule}) implies that, given an equal number of lanes,
green times are more likely for long queues,
which could be said to exert some ``pressure'' on the traffic light. However, if
road sections $j$ demanded by turning flows are congested, this exerts some ``counter-pressure''.
This will suppress green lights in cases where they would not allow to serve vehicles,
i.e.\ where they would not make sense. As a consequence, while cycle times increase with growing arrival
rates as long as these can be served,  they may go down
again when the road network is too congested.

\subsection{Oscillations at a merge bottleneck}

For the purpose of illustration,
let us discuss a merge bottleneck (see Fig.~\ref{fig:scenario}).
The two merging road sections $i\in \{1,2\}$ shall have
the overall capacities $I_i Q^{\rm max}$ with $I_1 \ge I_2$, while the subsequent section $j$ shall have
the capacity $I_j Q^{\rm max}\ge I_1 Q^{\rm max}$, so that no congestion will occur
in the subsequent road section. Let us assume that the arrival flows $Q_i^{\rm arr}$ are constant
in time.
Furthermore, let us assume that the traffic light for road section 2 turns red at
times $t_0$, $t_2$, etc., while the red lights for road section 1 start at $t_1$, $t_3$, etc.
The green times for road section 1 begin after an yellow time period of $\tau$, i.e.\ at
times $t'_{2k} = t_{2k} +\tau$ and last for the time periods $t_{2k+1} - t'_{2k}$.
\par
\begin{figure}[htbp]
    \begin{center}
    \includegraphics[width=1\textwidth]{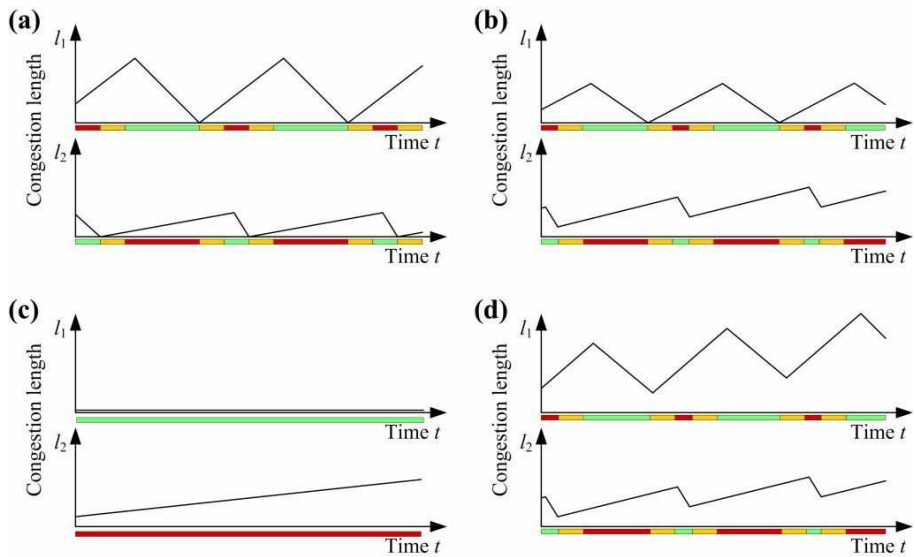}
    \caption{Different cases of the self-organized control of a merge bottleneck:
(a) The vehicle queue in each road section is completely
cleared, before the traffic light turns red. (b) The traffic light in the side road turns red, before the vehicle queue
has fully disappeared, but the main road is fully cleared.
(c) In extreme cases, if a maximum cycle time is not enforced,
the side road would never get a green light and the main road would always be served.
(d) When the sum of arrival rates is higher than the capacity of the
subsequent road section, the vehicle queues in both
road sections may grow under certain conditions (see text).}
    \label{fig:modi}
    \end{center}
\end{figure}

We can distinguish the following cases:
\begin{itemize}
\item[1.] Equivalent road sections:
If $I_1 = I_2$, the queues on both road sections will be completely cleared in an
alternating way, see Fig.~\ref{fig:modi}(a).
In case of growing vehicle queues, the green times grow accordingly.
\item[2.] One main and one side road ($I_1> I_2$):
\begin{itemize}
\item[(i)] If the arrival flow $Q_2^{\rm arr}$ of road section 2 (the side road) is low, both roads
are completely cleared.
\item[(ii)] In many cases, however, the
queue length in the side road grows in the course of time, while
the queue in the main road (road section 1) is completely cleared, see Fig.~\ref{fig:modi}(b).
As a consequence, road section 2 will be fully congested after some time period,
which limits a further growth of the queue and discourages drivers to use this
road section according to our traffic assignment rule. In extreme cases, when no
maximum cycle time is implemented (see Sec.~\ref{restred}),
the main road may have a green light all the time,
while road section 2 (the side road) is never served, see Fig.~\ref{fig:modi}(c).
\item[(iii)] If the sum $\sum_i I_i Q_i^{\rm arr}$ of overall arrival flows exceeds
the capacity $I_j Q^{\rm max}$ of the subsequent road section $j$, the queue on
both road sections will grow, see Fig.~\ref{fig:modi}(d).
\end{itemize}
\end{itemize}
We will now discuss these cases in more detail.

\subsubsection{Equivalent road sections}
Let us assume the queue length on road section 2 is zero at time $t_0$ and the
traffic light switches to red in order to offer a green light to road section 1 at time
$t'_0 = t_0 + \tau$. The queue length at time $t$ is given by
\begin{equation}
 l_1(t) = l_1(t'_0) + C_1 (t - t'_0) \, ,
\end{equation}
where
\begin{equation}
 C_i = \left( \frac{\rho^{\rm jam}}{Q_i^{\rm arr}} - \frac{1}{V_i^0}\right)^{-1}
 = \frac{Q_i^{\rm arr}}{\rho^{\rm jam} - Q_i^{\rm arr}/V_i^0}
\end{equation}
according to Eq.~(\ref{spe}). Note that, in the limit of small arrival rates $Q_i^{\rm arr}$, this
queue expansion velocity is proportional to $Q_i^{\rm arr}$. The reduction of the
queue starts with the green phase and is proportional to $c$. We, therefore, have the
following equation for the length of the effective queue (= queue length minus
area of quasi-free traffic):
\begin{equation}
 l_1^{\rm eff}(t) = l_1(t) + c(t-t'_0)  = l_1(t'_0) + C_1(t-t'_0) - |c| (t-t'_0) \, .
\end{equation}
The effective queue length disappears at time
\begin{equation}
 t_0^* = t'_0 + \frac{l_1(t'_0)}{|c| - C_1} \, .
\end{equation}
However, the last vehicle of the queue needs an additional time period of $l_1(t_0^*)/V_1^0$
to leave the road section, so that the queue length $l_1(t)$ in road section 1 becomes zero
at time $t=t_1$ with
\begin{equation}
 t_1 = t_0^* + \frac{l_1(t_0^*)}{V_1^0} = \dots = t'_0 + l_1(t'_0) \frac{1+|c|/V_1^0}{|c| - C_1} \, .
\label{one}
\end{equation}
At that time, the traffic light for road section 1 switches to red and road section 2 is served by
a green light starting at $t'_1 = t_1 + \tau$. Analogous considerations show that
the queue in road section 2 is cleared at time
\begin{equation}
  t_2 = t'_1 + l_2(t'_1) \frac{1+|c|/V_2^0}{|c| - C_2} \, .
\end{equation}
The next green time for road section 1 starts at time $t'_2 = t_2 + \tau$ and ends at
\begin{equation}
  t_3 = t'_2 + l_1(t'_2) \frac{1+|c|/V_1^0}{|c| - C_1} \, .
\label{three}
\end{equation}
We can determine the queue length $l_1(t'_2)$ at the beginning of the green phase as the
queue length that has built up during the previous red phase of length $t_2 - t'_1$
and two yellow phases of duration $\tau$ each. As a consequence, we find
$l_1(t'_2) = C_1(t_2 - t'_1 + 2\tau)$. In the stationary case we have $l_1(t'_2)
= l_1(t'_0)$ and $l_1(t_1) = 0$, as the queue on road section 1 is completely cleared
at time $t_1$.  This eventually leads to a rather complicated formula for $t_2 - t'_1$,
which is proportional to the respective queue length. For small values of
the arrival rates $Q_i^{\rm arr}$, one can show that the green times are proportional
to $C_i$ and $Q_i^{\rm arr}$. That is, the duration of the green phases is proportional to the
arrival rates, as expected, if the arrival rates are small enough. The cycle time grows linearly
with $Q_1^{\rm arr} + Q_2^{\rm arr}$.

\subsubsection{One main and one side road}\label{mainroad}
If both road sections are completely cleared as in case (i) above, the mathematical
treatment is analogous to the previous section. More interesting is case (ii), in which
the traffic light for road section 2 switches to red already before the queue is cleared
completely, see Fig.~\ref{fig:modi}(b). While Eqs.~(\ref{one}) and (\ref{three}) are still valid, we have to find
other expressions for $t_2$ and $l_1(t'_2) = l_1(t'_0)$. Let $t_2^+$ be the time point
in which the queue of length $l_1(t_2)$ in road section 1 at time $t_2$ would be
completely resolved, if the traffic light would turn green for road section 1 at time $t_2$.
Road section $1$ could for sure deliver an overall flow of $I_1 Q^{\rm max}$ between
$t'_2=t_2+\tau$ and $t_2^+$, while the departure flow from road section 1 could be much
smaller than $I_1Q^{\rm max}$ afterwards. In order to switch to green in favor
of road section 1, it is, therefore, reasonable to demand
\begin{equation}
 I_1 Q^{\rm max} [t_2^+ - (t_2+\tau)] \ge I_2 Q^{\rm max} (t_2^+ - t_2) \, .
\label{EQ}
\end{equation}
This formula considers the time loss $\tau$ by switching due to the intermediate yellow period,
and it presupposes that $Q^{\rm max}(t_2^+ - t_2) \ge l_2(t_2) \rho^{\rm jam}$, i.e.\ road
section 2 can maintain the maximum flow $Q^{\rm max}$ until $t_2^+$. Our philosophy is to give a
green light to the road section which can serve most vehicles during the next time period $t_2^+ - t_2$.
The equation to determine $t_2^+ = t_2^- + l_1(t_2)/V_1^0$ is
$l_1(t_2) = |c| [t_2^- - (t_2 + \tau)]$ with $l_1(t_2) = C_1 (t_2 - t_1)$. This leads to
$t_2^- = t_2 + \tau + C_1(t_2 - t_1)/|c|$ and
\begin{equation}
 t_2^+ = (t_2 + \tau) + \left( \frac{C_1}{|c|} + \frac{C_1}{V_1^0} \right) (t_2 - t_1) \, ,
\label{with}
\end{equation}
while Eq.~(\ref{EQ}) implies
\begin{equation}
 t_2^+ - t_2 \ge \frac{\tau}{1 - I_2/I_1} \, .
\end{equation}
Together with Eq.~(\ref{with}) we find
\begin{equation}
 t_2 - t_1 = \left. \frac{\tau}{I_1/I_2 - 1} \right/ \! \left( \frac{C_1}{|c|} + \frac{C_1}{V_1^0} \right) \, .
\end{equation}
For $I_1 = I_2$, one can immediately see that the traffic light would never switch before the queue
in road section 2 is fully resolved. However, early switching could occur for $I_1>I_2$.
\par
Once the traffic light is turned green at time $t_2$, the vehicles which have queued up until time $t_2^+$
will be served with the overall rate $I_1Q^{\rm max}$
as well, until the departure flow is given by the lower arrival flow $Q_1^{\rm arr}$ at time
$t_3$ and later. The time point $t_2^*$ at which the effective queue resolves is
given by $l_1(t_2^*) = |c|[t_2^* - (t_2 + \tau)]$, which results in
\begin{equation}
 t_2^* - t_2 = \frac{t_2^- - t_2}{1 - C_1/|c|} = \frac{\tau + C_1(t_2-t_1)/|c|}{1-C_1/|c|} \, .
\end{equation}
The last vehicle of the queue has left road section 1 at time $t_3$ with
\begin{equation}
 t_3 - t_2 = \frac{t_2^+ - t_2}{1 - C_1/|c|}
 = \frac{\tau}{(1-I_2/I_1)(1-C_1/|c|)}\, .
\end{equation}
Afterwards, the overall departure flow drops indeed to $I_1Q_1^{\rm arr}$, and the
traffic light tends to turn red if $I_1Q_1^{\rm arr} < I_2 Q^{\rm max}$. Otherwise, it
will continue to stay green during the whole rush hour. Considering $l_1(t_3) = 0
= l_1(t_1)$ and $l_1(t'_2) = C_1 (t'_2 - t_1)$, one can determine all quantities. One can show that
the green time fraction for road section 1 grows proportionally to $Q_1^{\rm arr}$, if $\tau$ is small.
Moreover, one can derive that the green time fractions of both road sections and the
cycle time $T^{\rm cyc} = t_3 - t_1$ are proportional to $C_1$, i.e.\ the main road dominates the
dynamics. The queue length on road section 2 tends to grow, as it is never fully cleared.
\par
If $I_1Q_1^{\rm arr} + I_2 Q_2^{\rm arr} > I_j Q^{\rm max}$, it can also happen that the queues
grow in both road sections. This is actually the case,
if $I_1Q_1^{\rm arr} > I_j Q^{\rm max}$, see Fig.~\ref{fig:modi}(d).
Moreover, in the case $I_2 Q^{\rm max} < I_1 Q_1^{\rm arr}$,
road section 2 would never be served, see Fig.~\ref{fig:modi}(c). This calls for one of several possible solutions:
1. Allow turning on red. 2. Decide to transform the side road into a dead end. 3. Build a
bridge or tunnel. 4. Use roundabouts or other
road network designs which do not require traffic lights.
5. Treat main and side roads equivalently, i.e.\ set $I_1 = I_2=1$
in the above formulas, or specify suitable parameter values for $I_i$, although it
will increase the overall delay times.
6. Restrict the red times to a maximum value at the cost of increased overall delay times
and reduced intersection throughput.

\subsubsection{Restricting red times}\label{restred}
In order to avoid excessive cycle times, one has to set upper bounds.
This may be done as follows: Let $T^{\rm max}$ be the maximum allowed cycle time,
\begin{equation}
 \overline{\gamma_i} = \frac{1}{T^{\rm max}}
\!\!\!\int\limits_{t-T^{\rm max}}^t \!\!\! dt'\; \gamma_i(t')
\end{equation}
the green time fraction within this time interval, and
\begin{equation}
 \overline{Q_i^{\rm arr}} = \frac{1}{T^{\rm max}}
\!\!\!\int\limits_{t-T^{\rm max}}^t \!\!\! dt'\; Q_i^{\rm arr}(t')
\end{equation}
the average arrival rate. If $\overline{\gamma_i}$ exceeds a specified green time fraction $u_i^0$,
the green light will be switched to red.
This approach also solves the problem that even small vehicle queues or single vehicles
must be served within some maximum time period.
\par
The green time fractions $u_i^0$ may slowly vary in time and could be specified proportionally to
the relative arrival rate $\overline{Q_i^{\rm arr}}/ \sum_{i'}\overline{Q_{i'}^{\rm arr}}$, with some
correction for the yellow time periods. However, it is better to determine
the green time fractions $u_i^0$ in a way that helps to optimize the system performance
(see Sec.~\ref{pricing}).

\subsubsection{Intersection capacity and throughput}
\begin{figure}[htbp]
    \begin{center}
        \includegraphics[width=1\textwidth]{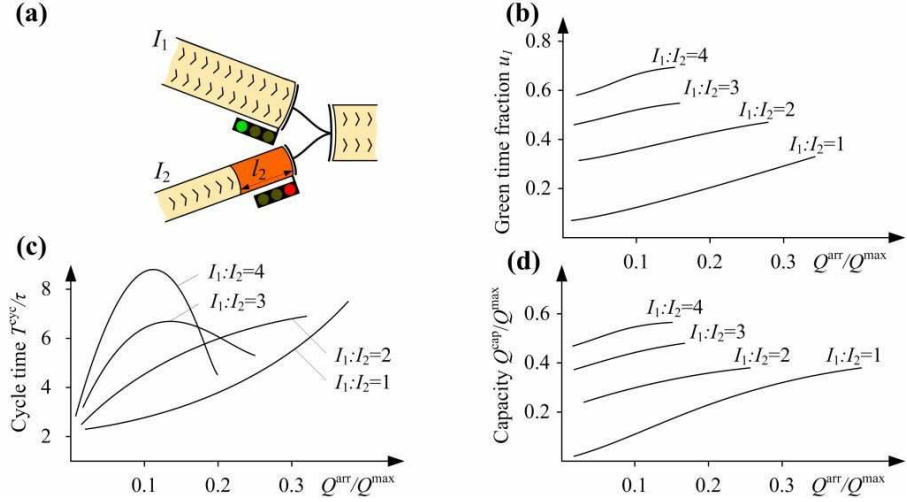}
    \end{center}
\caption{(a) Illustration of the traffic control of a merge bottleneck
for constant arrival rates and a non-congested outflow. The characteristic behavior of
the proposed self-organized traffic light control depends on the
number $I_i$ of lanes of the entering road sections $i$ and on the arrival rates $Q_i^{\rm arr}$:
(b) Actual green time fraction $u_1$ for $Q_2^{\rm arr}=\mbox{const.}$ and variable
$Q_1^{\rm arr}$, (c) cycle time $T^{\rm cyc}$ as compared to the yellow time period $\tau$
for $Q_2^{\rm arr} = Q_1^{\rm arr}$,
and (d) actual throughput $Q^{\rm all}$ of the signalized intersection in comparison
with the maximum uninterrupted flow $Q^{\rm max}$ per lane for $Q_2^{\rm arr} = Q_1^{\rm arr}$.}
\label{fig:scenario}
\end{figure}

Let us finally calculate the average
throughput $Q^{\rm all}$ of the signalized intersection. When the traffic
volume is low, it is determined by the sum $\sum_i I_i \overline{Q_i^{\rm arr}}$ of
average arrival flows, while at high traffic volumes, it is given by the intersection capacity
\begin{equation}
Q^{\rm cap} = Q^{\rm max} \frac{ I_1 (t_3-t'_2) + I_2 (t_2-t'_1)}{t_3-t_1}
= Q^{\rm max} \frac{ I_1 (t_3-t'_2) + I_2 (t_2-t'_1)}{T^{\rm cyc}} \, .
\end{equation}
This implies
\begin{equation}
 Q^{\rm all} = \min\left( \sum_i I_i \overline{Q_i^{\rm arr}}, Q^{\rm cap} \right) \, .
\end{equation}
According to these formulas, the losses in throughput and capacity
by the yellow times $2\tau$ are reduced by longer green times
$t_3-t'_2$ and $t_2-t'_1$. Our calculations indicate that our switching rule automatically increases
the cycle time $T^{\rm cyc} =t_3-t_1$ and the intersection capacity $Q^{\rm cap}$,
when the arrival rates $Q_i^{\rm arr}$ of equivalent roads with $I_1 = I_2$ or the
arrival rate $Q_1^{\rm arr}$ of a main road are
increased. Figure~\ref{fig:scenario} shows the cycle time $T^{\rm cyc}$, throughput $Q^{\rm all}$,
and green time fraction $u_1$ as a function of $Q_i^{\rm arr}=Q^{\rm arr}$ for different
values of $I_1/I_2$.
\pagebreak

\subsection{Serving single vehicles at low traffic volumes}\label{single}

While traffic lights have been invented to efficiently coordinate and serve vehicle flows
at high traffic volumes, they should ideally provide a green light for every arriving vehicle at
low average arrival rates $\overline{Q}_i^{\rm arr}$. According to formula (\ref{conds}),
the departure flow $Q_i^{\rm dep}(t)$ will, in fact, be 0 most of the time on all road sections. Only
during short time periods, single vehicles will randomly
cause positive values of $Q_i^{\rm arr}(t-L_i/V_i^0)$ on one of the road sections $i$.
The traffic light should be turned green shortly before the arrival of the vehicle at the
downstream boundary of this road section. If switching requires a time period of $\tau$,
the arrival flow $Q_i^{\rm arr}(t-L_i/V_i^0+\tau)$ would need to trigger a switching of the
traffic light in favor of road section $i$. Considering this and formula (\ref{conds}),
it is essential to take a switching decision based on the departure flow $Q_i^{\rm dep}(t+\tau)$
expected at time $t+\tau$. The departure flow $Q_i^{\rm dep}(t)$ can, in fact,
be forecasted for a certain time period based on available
flow data and assumed states of neighboring signals.
In order to minimize the time period $\tau$, it makes sense to
switch any traffic light to red, if no other vehicle is following. That is, at low traffic volumes,
all traffic lights would be red most of the time. However, any single vehicle would
trigger an anticipative green light upon arrival, so that vehicles would basically never have to
wait at a red light.

\subsection{Emergence of green waves through self-organized synchronization}\label{green}

\label{sec:Convoys} In order to let green waves emerge in a
self-organized way, the control strategy must show a tendency to
form vehicle groups, i.e.\ convoys, and to serve them just as they
approach an intersection. For this to happen, small vehicle clusters must
potentially be delayed, which gives them a chance to grow. When they are released, the
corresponding ``convoys'' may themselves trigger a green wave.

In fact, the ideal situation would be that traffic flow from road section $i$
arrives at location $L_j-l_j(t)$ in a subsequent road section $j$ just
when the effective queue $l_j^{\rm eff}(t)$ has resolved. This is equivalent
with the need to arrive at location $L_j$ just at the moment
when the queue length $l_j(t)$ becomes zero. Under such conditions, free arrival flows
$Q_j^{\rm arr}(t-L_j/V_j^0)$ with values around $Q_i^{\rm max}$
would immediately follow the high outflow $Q_j^{\rm dep}= Q_j^{\rm max}$ from
the (resolving) congested area in road section $j$ (here, we assume $I_i = I_j$).
As a consequence, the green light at the end of road section
$j$ would be likely to continue. This mechanism could establish a synchronization among traffic lights,
i.e.\ a green wave by suitable adjustment of the time offsets, triggered by vehicle flows. As it requires a time
period $\Delta t_j = [L_j-l_j(t)]/V_j^0$ to reach the upstream congestion front in section $j$, it will be required
to turn the signal of the previous road section $i$ green a time period $\Delta t_j$ before
the effective queue is expected to resolve. This time period defines the necessary forecast time interval.
\par
When the effective queue of length $l_j^{\rm eff}(t)$ is resolved, the
related sudden increase in $L_j-l_j(t)$ can cause a sudden increase in
$\Delta N_i^{\rm pot}$ and, thereby, possibly trigger a switching of the traffic light.
The emergence of green waves obviously requires
that the green light at the end of road section $j$ should stay long
enough to resolve the queue. This is likely, if road section $j$ is a main road (arterial),
see our considerations in Sec.~\ref{mainroad}.
\par
In a more abstract sense, the intersections in the road network can be understood as
self-sustained oscillators which are coupled by the vehicle flows
between them. Therefore, one might expect them to synchronize like
many natural systems do \cite{Synchronization}.
Interestingly, even if the intersections are not coupled
artificially with some communication feedback, the weak coupling
via vehicle flows is sufficient to let larger areas of the road
network synchronize. The serving direction percolates through the
network, stabilizes itself for a while and is then taken over by
another serving direction.
In other words, neighboring intersections affect each other by interactions via vehicle flows, which
favors a mutual adjustment of their rhythms. This
intrinsic mechanism introduces order, so that vehicle flows are coordinated.

\section{Summary and outlook} \label{summa}

In this contribution, we have presented a section-based traffic model for the
simulation and analysis of network traffic. Moreover, we have proposed
a decentralized control strategy for traffic flows, which has certain interesting features:
Single arriving vehicles always get a green light. When the
intersection is busy, vehicles are clustered, resulting in an
oscillatory and efficient service (even of intersecting main flows).
If possible, vehicles are kept going in order
to avoid capacity losses produced by stopped
vehicles. This principle
bundles flows, thereby generating main flows
(arterials) and subordinate flows (side roads and residential areas). If a
road section cannot be used due to a building site or an accident,
traffic flexibly re-organizes itself. The same applies to
different demand patterns in cases of mass events, evacuation
scenarios, etc. Finally, a local dysfunction of sensors or control
elements can be handled and does not affect the overall system. A
large-scale harmonization of traffic lights is reached by a
feedback between neighboring traffic lights based on the vehicle flows themselves,
which can synchronize traffic signals and organize green waves.
In summary, the system is self-organized
based on local information, local interactions, and local
processing, i.e.\ decentralized control. However, a multi-hierarchical
feedback may further enhance system performance by increasing the
speed of large-scale information exchange and the speed of
synchronization in the system.
\par
We should point out some interesting differences compared to conventional
traffic control:
\begin{itemize}
\item The green phases of a traffic light depend on the respective traffic
situation on the previous {\em and} the subsequent road sections. They are
basically determined by actual and expected queue lengths and delay times.
If no more vehicles need to be
served or one of the subsequent road sections is full, green times
for one direction will be terminated in favor of green times for other directions.
The default setting corresponds to red lights, as this enables one to respond quickly to
approaching traffic. Therefore, during light traffic
conditions, single vehicles can trigger a green light upon arrival at the traffic signal.
\item Our approach does not use precalculated or predetermined signal plans.
It is rather based on self-organized red and green phases. In particularly, there is
no fixed cycle time or a given order of green phases. Some roads may be even served
more frequently than others. For example, at very low traffic volumes it can
make sense to  serve the same road again before all other road sections have been served.
In other words, traffic optimization is not just a matter of green times and their permutation.
\item Instead of a traffic control center, we suggest a distributed, local control
in favor of greater flexibility and robustness. The required information can
be gathered by optical or infrared sensors, which will be cheaply available
in the future. Complementary information can be obtained by a coupling with
simulation models. Apart from the section-based model proposed in this paper,
one can also use other (e.g.\ microsimulation) models with or without
stochasticity, as our control approach does not depend on the traffic model.
Travel time information to enhance route choice decisions may
be transmitted by mobile communication.
\item Pedestrians could be detected by modern sensors as well and handled
as additional traffic streams. Alternatively, they may get green times during compatible
green phases for vehicles or after the maximum cycle time $T^{\rm max}$.
Public transport (e.g.\ busses or trams) may be treated as vehicles with a higher weight.
A natural choice for the weight would be the average number of passengers.
This would tend to prioritize public transport and to give it a green light upon arrival
at an intersection. In fact, a prioritization of public transport harmonizes much better
with our self-organized traffic control concept than with precalculated signal plans.
\end{itemize}

\subsection{Future research directions}

\subsubsection{Towards the system optimum}\label{pricing}

Traffic flow optimization in networks is not just a matter of durations, frequencies,
time offsets and the order of green times, which may be adjusted in the way described above.
Conflicts of flows and related inefficiencies can also be a result of the following problems:
\begin{itemize}
\item Space which is urgently required for certain origin-destination flows may
be blocked by other flows, causing a spill-over and blockage of upstream road
sections. One of the reasons for this is the cascaded minimum function (\ref{eq:DivergingNode}).
It may, therefore, be helpful to restrict turning only to subsequent road sections
that are normally not fully congested (i.e.\ wide and/or long road sections).
\item Giving green times to compatible vehicle flows may cause the over-proportional
service of certain road sections. These over-proportional flows may be called parasitic.
They may cause the blockage of space in subsequent road sections which would
be needed for other flow directions. In order to avoid parasitic flows, it may be useful
to restrict the green times of compatible flow directions.
\item Due to the selfish route choice behavior, drivers tend to distribute over alternative routes
in a way that establishes a Wardrop equilibrium (also called a Nash or user equilibrium)
\cite{Papa91}. This reflects the tendency of humans to balance travel times \cite{experiment}. That is, all
subsequent road sections $j$ of $i$ used to reach a destination $d$ are characterized by
(more or less) equal travel times. If the travel time on one path was less than on alternative ones, more
vehicles would choose it, which would cause more congestion and a corresponding increase
in travel times.
\end{itemize}
In order to reach the {\em system optimum}, which is
typically defined by the minimum of the overall travel times,
the drivers have to be coordinated. This would be able to further enhance the capacity
of the traffic network, but it would require the local adaptation of signal control parameters.
For example, the enforcement of {\em optimal} green time fractions $u_i^0$ based on the method
described in Sec.~\ref{restred} would be one step into this direction, as it is not necessarily the best, when
green time fractions are specified proportionally to the arrival rates $Q_i^{\rm arr}$.
\par
Unfortunately, green time fractions $u_i^0$ do not allow to differentiate between different origin-destination
flows using the same road section. Such a differentiation would allow one
to reserve certain capacities (i.e.\ certain fractions of road sections) for specific flows. This
could be reached by advanced traveller information systems (ATIS) \cite{Hu,MahJou,selten}
together with suitable pricing schemes, which would increase the
attractiveness of some routes compared to others.
\par
Different road pricing schemes have been proposed,
each of which has its own advantages
and disadvantages or side effects. Congestion charges, for example,
could discourage to take congested routes required to reach
minimum {\em average} travel times, while conventional tolls and road pricing may reduce the trip
frequency due to budget constraints (which potentially interferes with economic growth and
fair chances for everyone's mobility).
\par
In order to activate capacity reserves, we therefore
propose an automated route guidance system based on the following principles:
After specification of their destination, drivers should get
individual route choice recommendations in agreement with the traffic situation and
the route choice proportions required to reach the system optimum.
If an individual selects a faster route instead of the recommeded route it should, on the one
hand, have to pay an amount proportional to the increase in the overall travel time compared
to the system optimum. On the other hand, drivers not in a hurry should be encouraged to take the
slower route $i$ by receiving the amount of money corresponding to the related decrease in
travel times. Altogether, such an ATIS could support the system optimum while allowing
for some flexibility in route choice. Moreover, the fair usage pattern would be cost-neutral
for everyone, i.e.\ traffic flows of potential economic relevance would not be suppressed by extra costs.

\subsubsection{On-line production scheduling} \label{product}

Our approach to self-organized traffic light control could be also transfered
to a flexible production scheduling, in order to cope with problems of multi-goal
optimization, with machine breakdowns, and variations
in the consumption rate. This could, for example, help to optimize the difficult problem of
re-entrant production in the semiconductor industry \cite{Kempf,Armbruster,TGF03}.
\par
In fact, the control of network traffic flows shares many features
with the optimization of production processes. For example, travel
times correspond to cycle times, cars with different origins and
destinations to different products, traffic lights to production
machines, road sections to buffers. Moreover, variations in
traffic flows correspond to variations in the consumption rate,
congested roads to full buffers, accidents to machine breakdowns,
and conflicting flows at intersections to conflicting goals in
production management. Finally, the cascaded minimum function
(\ref{eq:DivergingNode}) reflects the fact that the scarcest
resource governs the maximum production speed: If a specific
required part is missing, a product cannot be completed. All of
this underlines the large degree of similarity between traffic
and production networks \cite{TGF03}. As a consequence, one can apply similar
methods of description and similar control approaches.

\begin{acknowledgments}
This research project has been partially supported by the German
Research Foundation (DFG project He 2789/5-1). S.L. thanks for his
scholarship by the ``Studienstiftung des Deutschen Volkes''.
\end{acknowledgments}

\end{document}